\documentclass[twoside]{article}
\usepackage{amsmath, amsfonts, amssymb}
\usepackage{aat}
\usepackage{aattable}
\allowdisplaybreaks

\begin{document}
\thispagestyle{myfirst}

\setcounter{page}{1}

\mylabel{19}{190}  
\mytitle{EVOLUTION OF ISOLATED NEUTRON STARS IN GLOBULAR CLUSTERS:
NUMBER OF ACCRETORS}
\myauthor{S.B. POPOV$^1$ and M.E. PROKHOROV$^1$}
\myadress{$^{\it 1}$Sternberg Astronomical Institute, Moscow, Russia,
119899, Universitetski pr. 13}
\mydate{(Received February 12, 2001)}
\myabstract{ With a simple model from the point of view of population
synthesis  we try to verify an interesting
suggestion made by Pfahl \& Rappaport (2001) that dim sources in globular
clusters (GCs) can be isolated accreting neutron stars (NSs).
Simple estimates show, that we can expect about 0.5-1 accreting isolated
NS per typical GC with $M=10^5 M_{\odot}$ in correspondence with
observations. Properties of old accreting isolated NSs in GCs
are briefly discussed. We suggest that accreting NSs in GCs experienced
significant magnetic field decay.}
\mykey{Neutron stars, globular clusters, accretion}

\section{INTRODUCTION}
Galactic isolated neutron stars (NSs) are expected to be as many as
$10^8$--$10^9$, a non--negligible fraction of the total
stellar content of the Galaxy. In the last 10 years they received
special attention. The idea of observing such objects in the X-ray range has
emerged in soon after discovery of the first X-ray sources
(Ostriker {\it et al.} 1970).
Now we are sure that few dim ROSAT sources are nearby
isolated neutron stars (see Neuh\"auser 2001, Walter 2001). 

All these sources emit a thermal spectrum at $\approx 100$ eV and the
derived column densities place them at relatively close distances ($\leq
150$ pc) with luminosities $L\sim 10^{30}$--$10^{31}$ erg $\rm s^{-1}$.
Several of them can be visible due to accretion
from the interstellar medium (ISM), others --- due to thermal emission of
cooling relatively young, $\sim 1$ Myr, NSs (see Treves {\it et al.} 2000 for
a recent review of accreting isolated NSs, and Popov {\it et al.} 2000b
for discussion on relative fraction of cooling and accreting NSs
in the solar vicinity).

 X-ray observations of globular clusters (GCs) show  large
population, $>30$, of dim X-ray sources (hereafter dim GC sources)
with $L\sim 10^{31}$--$10^{34}$ erg $\rm s^{-1}$ (Verbunt 2001).
Pfahl \& Rappaport (2001) suggested, that part of dim GC sources
can be old isolated accreting NSs. In this short note we try
to analyze this idea from the point of view of population synthesis
(see Lipunov {\it et al.} 1996 for detailed description of population synthesis
method).

 Population synthesis of isolated NSs in the Galactic disk was made
previously by several authors 
(see Popov {\it et al.} 2000a,b for recent results and references).
Evolution of isolated NSs in the Galactic disk and in GCs
must be significantly different.
Here we try to explore with a simple model
evolution of progenitors of isolated accreting NSs in GCs.
In the next section we briefly discuss our models. In section 3 we present
our results, and in the last section we discuss properties of old accreting
NSs in GCs.

\section{THE MODEL}
 An isolated NS can pass through four main stages of evolution:
Ejector, Propeller, Accretor and Georotator (see Lipunov 1992 for
detailed description of each stage).
We consider, that all NSs are born as Ejectors, and then
while they slow down (or/and their magnetic field decays) they
pass through the Propeller and Accretor stages. If spatial velocity
or/and magnetic field of a NS are very high, then this NS can
appear as Georotator after the end of the stage of ejection.

 For constant magnetic field
we can estimate the period of time, which a NS spends on the stage of
Ejector, $t_E$, as (see for example Popov \& Prokhorov 2000):

\begin{equation}
t_E\sim 10^{9} \, {\rm yrs}\, \mu_{30}^{-1} n^{-1/2} v_{10},
\end{equation}
where $\mu_{30}=\mu/10^{30}$ G cm$^3$ --- magnetic moment,
$v_{10}=v/10$ km s$^{-1}$ --- velocity relative to ISM, 
$n$ --- ISM number density.

For the Propeller stage one can find in the literature
a lot of different spin-down regimes,
but in the case of constant magnetic field 
the Propeller stage is always
considered to be much shorter than the Ejector stage
(see Lipunov \& Popov 1995). Taking this into account we neglect the time
spent by NSs on the Propeller stage in our simple estimates (for field decay this
stage become very important! see Livio {\it et al.} 1998, Colpi {\it et al.} 1998, 
Popov {\it et al.} 2000a).

 We calculate relative fraction of Accretors  
(i.e. objects with $t_E<t_{Hubble}=10^{10} \, {\rm yrs}$ which cannot appear as
Georotators due to relatively low
magnetic field or/and spatial velocity) 
among other isolated NSs for typical disk and GC
conditions after $10^{10}$ yrs of evolution. 
In the case of the Galactic disk we use constant starformation rate,
and in the case of GC starformation rate is considered as a
$\delta$-function (i.e. a starformation burst at $t=0$).
 
 For both cases we use Maxwellian velocity distribution
with $\sigma=140$ km s$^{-1}$, and log-normal distribution for magnetic
fields of NSs typical for radiopulsars (see Popov {\it et al.} 2000a,b for
discussion of these choices). 
For GC velocity was truncated at 30 km s$^{-1}$, which is a typical 
escaping velocity from the cluster.
For the Galactic disk we use typical value of ISM number density
$n=1$ cm$^{-3}$, for GCs --- $n=100$ cm$^{-3}$ 
(see Pfahl \& Rappaport 2001 and short discussion below).
In our simple model $n$, $\mu$ and $v$ are considered to be constant
during the whole evolution.

 For both populations (disk and GCs NSs) we assume the same parameters
of NSs (initial spin periods, moments of inertia, masses etc.).
Initial spin periods are assumed to be equal to 0.020 sec for all NSs
in our calculations. Our previous estimates (Popov {\it et al.} 2000a) show
that small variations of this parameter do not change final results for
Accretors.

\section{CALCULATIONS AND RESULTS}
 Our aim is to calculate the 
number of accreting isolated NSs per typical GC
with the total mass $10^5 M_{\odot}$ (we neglect the fact, that the GC
mass slightly decreases during GC evolution).

As the first step we compare fractions of Accretors in GCs and in the
Galactic disk. To do it we just run our simple models (described in the
previous section) for the same numbers of NSs for each case.
An important result is the following: for GCs fraction of Accretors
relative to other stages
is about 26 times lower than in the Galactic disk. 
Of course it is just a rough estimate,
but it shows, that even for high ISM density 
($n= 100 $ cm$^{-3}$) due to escaping of 
the most part of population of
isolated NSs  number of isolated accreting NSs in GCs is small.
Escaped NSs are considered as Ejectors as far as they have high spatial
velocities and move in low-density surrounding. One can compare typical
escaping velocity, 30 km s$^{-1}$, with typical velocity of Accretors
in the Galactic disk (Popov {\it et al.} 2000b), $>50$ km s$^{-1}$.  

 Then we try to estimate the number of Accretors for a typical GC.
In the Galactic disk Accretors form about 1\% of the total NS population
(Popov {\it et al.} 2000a).
Mass of the disk population of stars
is about $7\cdot 10^{10} M_{\odot}$.
Typical mass of a GC is assumed to be $10^5 \, M_{\odot}$. 
We note, that for the Galactic disk
fraction of 1\% was calculated (Popov {\it et al.} 2000a) for constant
starformation rate, for GCs  we calculate it for
a starformation burst (age of the burst is equal to the age of the disk).

In the disk we expect about $N=10^9$ isolated NS. I.e. about $10^7$ 
Accretors. After simple calculations we obtain about 
$(10^9\cdot 0.01 \cdot 10^{5})/(26\cdot 7\, \cdot 10^{10})\approx 0.55$ 
Accretors per
typical GC ($M=10^5 M_{\odot}$). If total number of NSs in the disk is
lower, than our estimate should be lower too.

We can also calculate the number of Accretors per GC in a more simple way.
Let us take the Salpeter mass function, and assume, that NSs are born
from mass interval 10--40 solar masses (Lipunov {\it et al.} 1996). 
In a simple model described above (starformation burst,
$10^{10}$ years of evolution, constant parameters of the ISM
and  NSs during their evolution)
we can calculate that about 0.25\%
NSs become Accretors in GCs (mainly because most part of NSs leave GCs
due to high kick velocity, but nearly all NSs which stay bounded with the GC
become Accretors). It gives us about 1.1 Accretor per GC.
It can be considered as an upper limit on the number of Accretors,
as far as for GCs number of massive stars can be lower
(initially) than it follows from the Salpeter mass function,
and evolution of NSs can proceeds with an average ISM density lower than
the value $n=100$ cm$^{-3}$ which we used in our calculations: all these
factors decrease the number of Accretors.

So, we can roughly estimate the number of Accretors per GC 
with the total mass $10^5$ solar masses as 0.5-1. We also note, that
coincidence of the two estimates can be considered as an additional (but not
very strong) argument for the number of Galactic NSs equal to $10^9$.

 To summaries, simple evolutionary estimates do not contradict 
the idea, that at least part of dim X-ray sources in GCs
can be old accreting isolated NSs. It is important to repeat such
calculations in more details (spin evolution,
dynamical evolution, realistic ISM distribution etc.) 
and with realistic models of field decay.
 
\section{DISCUSSION}
  It is reasonable to discuss in brief the following important topics:

\noindent
1. Interstellar gas density in GCs.

\noindent
2. Observed temperatures of dim sources in GCs.

\noindent
3. Periods of accreting NSs in GCs
(no pulsations were detected from dim GC sources)
and rotational equilibrium (period changes).

\noindent
4. Magnetic field distribution for Accretors in GCs.

 Pfahl \& Rappaport in their paper suggested and used high value of the
gas number density in GCs: $100 $ cm$^{-3}$. We used the same value in our
calculations. Actually, nobody observed gas in GCs at such high density.  
But it is reasonable to expect these high values and observations
provide high upper limits: $\sim (50$--$100)$ cm$^{-3}$ (Knapp {\it et al.} 1996).

 Average density along the track during a NS evolution can be lower than
the high value we used. This effect will increase the duration of the
Ejector stage (see eq.1), and correspondently decrease the number of Accretors.
But if we {\it a priori} consider dim X-ray sources to be Accretors,
we have to take high number density in order to explain observed
luminosities (Pfahl \& Rappaport 2001). The same high density comes from
estimates of gas accumulation in GCs due to mass loss from the stars
(Knapp {\it et al.} 1996, Pfahl \& Rappaport 2001).

That's why we argue, that the 
effect of decreasing of the number of Accretors is not very significant. Otherwise,
in the case of low density the hypothesis of accreting isolated NSs
should be rejected due to difficulties with the explanation of X-ray
luminosity of the sources under discussion.

 Relatively high temperatures of dim GC sources (0.1-0.5 keV),
which is higher than for dim ROSAT INS candidates 
($<200$ eV with typical values 50-100 eV, 
see Treves {\it et al.} 2000, Walter 2001, Neuh\"auser 2001) 
are in favour of the accreting hypothesis (both for single and binary NSs).
For example, young cooling NSs, age $\sim 1$ Myr, 
should be cooler than Accretors in the case of polar cap accretion.  
In Popov{\it  et al.} 2000b the authors plot temperature distribution
for Accretors (without magnetic field decay) for the disk population.
Maximum of the distribution corresponds to $\sim 400$--$500$ eV,
the total range is about 100--1000 eV (depending on the magnetic field and
accretion rate). But as far as luminosities of dim GC sources are
higher than luminosities of the disk accreting sources
(the ISM density is higher), they should
be also slightly warmer. May be accretion in that sources proceeds onto very
large polar cap due to magnetic field decay (objects are very old).

 Old Accretors in disk should have relatively long periods if 
their magnetic field
does not decay significantly (Lipunov \& Popov 1995). 
But in GCs spin periods for the same magnetic field
should be much shorter,
as far as the ISM density is higher:

\begin{equation}
p_A\sim 42 \, \mu_{30}^{6/7}v_{10}^{9/7}n_{100}^{-3/7} \, {\rm s}.
\end{equation}
Here $p_A$ is a critical period for transition from the stage of Propeller
to the stage of accretion (Lipunov 1992).
On that stage a NS continues to spin down till it comes to rough equilibrium
with turbulized ISM. So, typical periods should be close to some value,

\begin{equation}
p_{turb}\sim (100-300) \, v_{10}^{13/3}n_{100}^{-2/3}\mu_{30}^{2/3}\,
{\rm s}, 
\end{equation}

\noindent
for $n=100$ cm$^{-3}$ (Lipunov \& Popov 1995,
Konenkov \& Popov 1997), slightly longer than $p_A$.
Realistic period distribution for accreting isolated NSs
will be presented elsewhere (Popov, Prokhorov \& Khoperskov 2001).
It is characterized by broad maximum near $p_{turb}$, slightly shifted
to longer periods. 

 In the picture described above spin periods of isolated accreting NSs in GCs 
can be observed, and lack of period detection
is not in favour of Pfahl \& Rappaport (2001) hypothesis. To solve this
problem it is necessary to introduce something non-standard for NSs in GCs. 
For example significant field decay, so that Alfven radius
will be about the size of a NS, and no pulsations can be observed in that
case due to low level of modulation. 
As we note above it can also explain relation between temperature
and luminosity of that sources (due to large polar cap area).

 Periods and their derivatives (if detected) 
should show significant fluctuations
on the time scale $R_G/v\sim v^{-3}\approx 5.9 \, {\rm yrs} \, v^{-3}_{10}$. 
On the same time scale one expects
variations of luminosity, and they are observed (Pfahl \& Rappaport 2001)
on the time scale $\sim 1$ year. Detection of spin periods and analysis
of their correlation with luminosity changes are very desirable.

 Magnetic fields of Accretors in GCs should be, on average, smaller
than in disk, because it is much easier to become Accretor in a GC 
(density is
higher, and typical spatial velocity is lower: in disk it is about 50 km s$^{-1}$,
and in GCs about 20 km s$^{-1}$). 
For lower magnetic fields it is more difficult to reach $p_A$ (see eq.2),
so there should be low values of the magnetic field 
for which only in high density environment of
GC it is possible for a NS to become an Accretor. 

 We also note, that field decay can both: decrease and increase number of
Accretors (Colpi {\it et al.} 1998, Livio {\it et al.} 1998, 
Popov {\it et al.} 2000a, Popov
\& Prokhorov 2000). So, significant field decay, which make it easier for a
NS to reach the stage of accretion, can compensate the effect of lower
average ISM density along the evolutionary path of the NS.

 In principle 
some of dim GC sources can also be Georotators accreting warm ISM 
(aka MAGACs, see Rutledge 2001). These objects should be 
relatively hard sources, without any periodicity on the timescale
$<10^5$ s. 
But for GC conditions very large magnetic moments
are required to become a Georotator, 
because in dense ISM ($n\sim 100$ cm$^{-3}$) among low
velocity isolated NSs, which can stay in the GC, only most magnetized
objects can become Georotators instead of becoming 
Propellers (and then --- Accretors).
It makes this hypothesis unlikely. If one try to explain dim GC sources
as young cooling NSs, then it is necessary to accept, that they are really
very young to be so hot and luminous. It is difficult to explain why 
in several GCs  young NSs are observed
(Andrei Zakharov noted to us, that there is a possibility of NS formation
in GCs due to recent star formation from the accumulated matter,
which was lost by cluster members on their late stages of evolution). 
Most probably dim GC sources are accreting isolated NSs 
with decayed magnetic fields in correspondence with Pfahl \& Rappaport (2001)
hypothesis. 

\subsection*{Acknowledgements}
We thank Andrei Zakharov for discussions and unknown referees for useful
comments.
This work was supported by the RFBR (01-02-06265), 
 and NTP ``Astronomy'' (1.4.4.1; 1.4.2.3) grants.

\subsection*{References}

\rf{Colpi, M., Turolla, R., Zane, S., and Treves, A. (1998)
{\it ApJ} {\bf 501}, 252.}
\rf{Knapp, G.R., Gunn, J.E., Bowers, P.F., and Vasques Poritz, J.F. (1996)
          {\it ApJ} {\bf 462}, 231.}
\rf{Konenkov, D.Yu., and Popov, S.B. (1997) {\it Astronomy Letters} {\bf 23},
498.}
\rf{Lipunov, V.M. (1992) {\it Astrophysics of Neutron Stars} 
Berlin: Springer \& Verlag.}
\rf{Lipunov, V.M., and Popov, S.B. (1995) {\it AZh} {\bf 71}, 711.}
\rf{Lipunov, V.M., Postnov, K.A. and Prokhorov, M.E. (1996)
{\it Astron. and Space Phys. Rev.} {\bf 9}, part 4.}
\rf{Livio, M., Xu, C., and Frank, J. (1998) {\it ApJ} {\bf 492}, 298.}
\rf{Neuh\"auser, R. (2001) {\it Astron. Nachr.} {\bf 322}, 3 (astro-ph/0102004).}
\rf{Ostriker, J.P., Rees, M.J. and Silk J. (1970)
 {\it Astrophys. Letters} {\bf 6}, 179.}
\rf{Pfahl and Rappaport (2001) {\it ApJ} {\bf 550}, 172 (astro-ph/0009212).}
\rf{Popov, S.B., Colpi, M.,
Treves, A., Turolla, R., Lipunov, V.M., and  Prokhorov, M.E. (2000a)
{\it ApJ} {\bf 530}, 896.}
\rf{Popov, S.B., Colpi, M.,
Prokhorov, M.E., Treves, A., and Turolla, R. (2000b) {\it ApJ} {\bf 544}, L53.}
\rf{Popov, S.B., and Prokhorov, M.E. (2000) {\it A\&A} {\bf 357}, 164.}
\rf{Popov, S.B., M.E., and Khoperskov, A.V. (2001) (in preparation).}
\rf{Rutledge, R.E. (2001) {\it ApJ} (in press), astro-ph/0101550.}
\rf{Treves, A., Turolla, R., Zane, S., and Colpi, M. (2000) {\it PASP} {\bf
112}, 297.}
\rf{Verbunt, F. (2001) {\it A\&A} {\bf 368}, 137 (astro-ph/0012261).}
\rf{Walter, F. (2001) {\it ApJ} {\bf 549}, 433 (astro-ph/0009031).} 
   
\end{document}